\documentclass[12pt,letterpaper]{article}

\usepackage{amsmath,amssymb,calc}
\usepackage{amsthm}
\usepackage{graphicx}
\usepackage[nosort]{cite}

\usepackage{hyperref}

\makeatletter
\renewcommand\section{\@startsection {section}{1}{\z@}%
                                   {-3.5ex \@plus -1ex \@minus -.2ex}
                                   {2.3ex \@plus.2ex}%
                                   {\normalfont\large\bfseries}}
\renewcommand\subsection{\@startsection{subsection}{2}{\z@}%
                                     {-3.25ex\@plus -1ex \@minus -.2ex}%
                                     {1.5ex \@plus .2ex}%
                                     {\normalfont\bfseries}}
\makeatother

\let\non\nonumber

\let\a=\alpha
\let\b=\beta

\let\k=\kappa
\let\l=\lambda
\let\r=\rho
\let\s=\sigma

\def\one{^{(1)}}

\newcommand{\bea}{\begin{eqnarray}}
\newcommand{\eea}{\end{eqnarray}}
\newcommand{\bean}{\begin{eqnarray*}}
\newcommand{\eean}{\end{eqnarray*}}
\newcommand{\be}{\begin{equation}}
\newcommand{\ee}{\end{equation}}
\newcommand{\ben}{\begin{equation*}}
\newcommand{\een}{\end{equation*}}
\def\ba#1\ea{\begin{align}#1\end{align}}
\def\ban#1\ean{\begin{align*}#1\end{align*}}
\newcommand{\bma}{\begin{pmatrix}}
\newcommand{\ema}{\end{pmatrix}}

\newcommand{\hlf}{\frac{1}{2}}

\newcommand{\Z}{{\mathbb Z}}

\newcommand{\cR}{{\cal R}}

\newcommand{\La}{\Lambda}
\newcommand{\la}{\lambda}

\newcommand{\G}{\Gamma}

\newcommand{\vp}{\varphi}

\newcommand{\m}{\mu}
\newcommand{\n}{\nu}

\newcommand{\f}{\psi}

\newcommand{\nab}{\nabla}

\newcommand{\C}[1]{$(\ref{#1})$}

\def\IZ{\relax\ifmmode\mathchoice
{\hbox{\cmss Z\kern-.4em Z}}{\hbox{\cmss Z\kern-.4em Z}}
{\lower.9pt\hbox{\cmsss Z\kern-.4em Z}} {\lower1.2pt\hbox{\cmsss
Z\kern-.4em Z}}\else{\cmss Z\kern-.4em Z}\fi}
\def\IR{\relax{\rm I\kern-.18em R}}

\def\one{{\hbox{ 1\kern-.8mm l}}}

\def\tr{{\rm tr\,}}

\newlength{\bredde}
\def\slash#1{\settowidth{\bredde}{$#1$}\ifmmode\,\raisebox{.15ex}{/}
\hspace*{-\bredde} #1\else$\,\raisebox{.15ex}{/}\hspace*{-\bredde}
#1$\fi}

\newsavebox{\zzzbar}
\sbox{\zzzbar}
  {\setlength{\unitlength}{0.9em}
  \begin{picture}(0.6,0.7)
  \thinlines
  \put(0,0){\line(1,0){0.6}}
  \put(0,0.75){\line(1,0){0.575}}
  \multiput(0,0)(0.0125,0.025){30}{\rule{0.3pt}{0.3pt}}
  \multiput(0.2,0)(0.0125,0.025){30}{\rule{0.3pt}{0.3pt}}
  \put(0,0.75){\line(0,-1){0.15}}
  \put(0.015,0.75){\line(0,-1){0.1}}
  \put(0.03,0.75){\line(0,-1){0.075}}
  \put(0.045,0.75){\line(0,-1){0.05}}
  \put(0.05,0.75){\line(0,-1){0.025}}
  \put(0.6,0){\line(0,1){0.15}}
  \put(0.585,0){\line(0,1){0.1}}
  \put(0.57,0){\line(0,1){0.075}}
  \put(0.555,0){\line(0,1){0.05}}
  \put(0.55,0){\line(0,1){0.025}}
  \end{picture}}

\def\Re{{\rm Re ~}}

\newcommand{\ena}{\end{eqnarray}}
\newcommand{\beqa}{\begin{eqnarray}}
\newcommand{\eeqa}{\end{eqnarray}}

\def\G{\Gamma}

\renewcommand{\b}{\beta}
\newcommand{\g}{\gamma}

\def\a{\alpha}
\def\b{\beta}

\def\f{\phi}
\def\g{\gamma}

\def\k{\kappa}
\def\l{\lambda}
\def\m{\mu}
\def\n{\nu}

\def\om{\omega}

\def\r{\rho}
\def\s{\sigma}

\def\G{\Gamma}

 \numberwithin{equation}{section}

\def\1{{(1)}}
\def\2{{(2)}}
\def\3{{(3)}}

\def\1{{\bf 1}}
\def\a{{\alpha}}

\begin{document}
\begin{titlepage}

\today
\begin{center}

 \hfill        

\vskip 2 cm {\Large \bf Gaugino Condensation and the Cosmological Constant}
\vskip 1.25 cm {\bf  Callum Quigley}\non\\
\vskip 0.2 cm
 {\it Department of Mathematical and Statistical Sciences,\\  University of Alberta, Edmonton, AB T6G 2G1, Canada}\non\\ \vskip 0.2cm
e-mail: \href{mailto:cquigley@ualberta.ca}{cquigley@ualberta.ca}

\end{center}
\vskip 2 cm

\begin{abstract}
\baselineskip=18pt
The existence of de Sitter solutions in string theory is strongly constrained by no-go theorems. We continue our investigation of corrections to the heterotic effective action, with the aim of either strengthening or evading the these constraints. We consider the combined effects of H-flux, gauge bundles, higher derivative corrections and gaugino condensation. The only consistent solutions we find with maximal symmetry in four dimensions are Minkowski spacetimes, ruling out both de Sitter and anti-de Sitter solutions constructed from these ingredients alone.

\end{abstract}

\end{titlepage}


\section{Introduction}

The cosmological constant problem remains one of the greatest mysteries in all of theoretical physics. This is very likely a reflection of our ignorance of how to properly formulate quantum gravity in (asymptotically) de Sitter spacetime. While string theory has proven to be remarkably successful for understanding spacetimes with $\La\leq0$, the explicit construction of de Sitter solutions is by now a notoriously difficult problem. The reason is quite simple: an accelerating spacetime must violate the strong energy condition (SEC), but supergravity in ten and eleven dimensions cannot do so and this property is inherited upon compactification to  lower dimensions~\cite{Gibbons:1984kp,deWit:1986xg,Maldacena:2000mw,Gibbons:2003gb}. This no-go theorem extends, at the two-derivative level, to stringy localized sources, including branes, anti-branes and even orientifold planes~\cite{Dasgupta:2014pma}.

Higher derivative interactions provide a potential source of SEC violation. Unfortunately, even the leading order corrections to supergravity, including all couplings to fluxes, remain largely unknown. One notable exception, however, is the heterotic string, where the complete set of higher derivative corrections are known up to $O(\a'^3)$~\cite{Bergshoeff:1989de}. We analyzed these effects in~\cite{Green:2011cn}, and found that they do not suffice to produce the SEC violations required for de Sitter spacetimes. In fact, it was later shown that both dS and AdS spacetimes are ruled out in four dimensions to all orders in a perturbative $\a'$ expansion~\cite{Gautason:2012tb}. Non-perturbatively, worldsheet instanton corrections can lead to AdS vacua~\cite{Curio:2005ew,Anguelova:2010qd,Cicoli:2013rwa},\footnote{Note that these AdS solutions also require a gaugino condensate to stabilize the dilaton. Otherwise, there is a run-away potential to zero coupling and $\La=0$.} but dS solutions can be ruled out by an exact worldsheet argument~\cite{Kutasov:2015eba}.

In order to evade the no-go theorem we must, therefore, investigate beyond string-tree level. In this note we will focus our attention on a well-known $g_s$ effect, namely gaugino condensation. Not only does this quantum phenomenon probe the heterotic theory beyond tree-level, but it also generates a potential for the dilaton~\cite{Dine:1985rz,Derendinger:1985kk}. A priori, this non-perturbatively generated potential could have a positive minimum value, which would invalidate one of the key assumptions underlying the no-go theorem. Nevertheless, we find that the effects of gaugino condensation do not affect previous results: both  AdS and dS spacetimes remain excluded in four-dimensional compactifications of the heterotic string, up to $O(\a'^3)$.

The exclusion of heterotic AdS$_4$ solutions was initially surprising, given the number of explicit examples constructed in the literature. A closer inspection reveals that most of these solutions lie outside the realm of our analysis, relying on: threshold corrections to the gauge coupling~\cite{Gukov:2003cy,Anderson:2011cza}, worldsheet instantons (discussed above)~\cite{Anguelova:2010qd}, or a combination of these~\cite{Curio:2005ew,Cicoli:2013rwa}, strong coupling effects~\cite{Buchbinder:2003pi,Becker:2004gw,Buchbinder:2004nt,Gray:2007qy}, or other non-perturbative dynamics~\cite{Serone:2007sv}. However, there appear to be a few puzzling exceptions~\cite{deCarlos:2005kh,Frey:2005zz,Held:2010az,Klaput:2012vv}, which are at odds with our findings, since they engineer AdS$_4$ solutions using a combination of H-flux, $\a'$ corrections and gaugino condensates. According to our analysis, these solutions should not be possible. We address this conflict, and also the apparent tension with type IIB duals, at the end of this note.

The rest of this article is organized as follows. In Section~\ref{sec:review}, we review the $\a'$-corrected heterotic effective action and explain our ansatz for the four-dimensional solutions, including the gaugino condensate. In Section~\ref{sec:result}, we demonstrate that our ansatz implies the vanishing of the cosmological constant to $O(\a'^3)$. We also show that more general forms of the condensate do not alter this result, however we show that threshold corrections fall outside of the generalizations we consider. A sketch of how our argument should extend to all orders in $\a'$ is also presented. We end with a discussion of our results in Section~\ref{sec:discussion}, and possible directions to explore in the future. Our conventions are summarized in Appendix~\ref{app:conventions}, and some details of our dimensional reduction procedure are relegated to Appendix~\ref{app:reduction}.

\section{From ten to four dimensions}\label{sec:review}

\subsection{The heterotic effective action}

Our starting point will be the low energy effective action of the heterotic string, including $\a'$ corrections up to quadratic order. The massless field content of the theory consists of a metric $g_{MN}$, dilaton $\f$, NS two-form $B_{MN}$ with curvature $H_{MNP}$, and a $\rm{Spin}(32)/\Z_2$ or $E_8\times E_8$ gauge field $A_M$ with curvature $F_{MN}$. The fermions of the theory are all of Majorana-Weyl type and they are a gravitino $\psi_M$, dilatino $\la$, and an adjoint valued gaugino $\chi$. The neutral fermions will not be needed, and will therefore be suppressed.

The complete effective action was worked out in~\cite{Bergshoeff:1989de}\ through $O(\a'^2)$. In string frame, and setting  $\k^2=1/2$ for convenience, that action is given by\footnote{Our normalizations are obtained from~\cite{Bergshoeff:1989de}\ as follows: $\f\rightarrow e^{2\f/3}$, $H\rightarrow-\frac{1}{3\sqrt{2}}H$, $\chi\rightarrow\sqrt{2}\chi$, and $\beta \rightarrow\frac{\a'}{4}$. }
\ba\label{eqn:10daction}
S= \int d^{10}x\sqrt{-g}e^{-2\f}&\left[\cR +4|\nab\phi|^2 - \hlf\left|T\right|^2 -\frac{\a'}{4}\left(\tr|F|^2-\tr|R_+|^2 +2\tr\bar\chi D\!\!\!\!/\,\chi\right)\right],
\ea
where $\cR$ is the Ricci scalar and
$
R_+^{AB} = d\om_+^{AB} +\om_+^{AC}\wedge\om_+^{CB}
$
is the curvature two-form of the spin connection with torsion,
\ba
\om_\pm^{AB}{}_M = \om^{AB}{}_M \pm \hlf H^{AB}{}_M +O(\a'^2),
\ea
with $\om$ the standard spin connection. We also define the three-form
\ba\label{eqn:Sigma}
T_{MNP} = H_{MNP} + \frac{\a'}{8} \tr \bar\chi \G_{MNP} \chi.
\ea
Our conventions are for summarized in Appendix~\ref{app:conventions}. In addition to the explicit factors of $\a'$ appearing in~\C{eqn:10daction}-\C{eqn:Sigma}, the only other corrections to this order are captured by
\ba\label{eqn:H}
H = dB +\frac{\a'}{4}\left[CS(\om_+)-CS(A)\right],
\ea
where $CS(A)= \tr(A\wedge dA+\tfrac{2}{3}A\wedge A\wedge A)$ is the Chern-Simons three-form, and a similar expression holds for $CS(\om_+)$. This of course leads to the well-known Bianchi identity:
\ba\label{eqn:bianchi}
dH = \frac{\a'}{4}\left[\tr\left(R_+\wedge R_+\right) -\tr \left(F\wedge F\right)\right],
\ea
which must be satisfied, in addition to the equations of motion derived from~\C{eqn:10daction}, by any solution of the theory. Corrections to~\C{eqn:10daction}\ begin at $O(\a'^3)$.

\subsection{Four-dimensional ansatz: bosons}\label{sss:ansatz}

We seek four-dimensional solutions of~\C{eqn:10daction}\ with maximal symmetry.
The most general ansatz we can make for the full ten-dimensional spacetime is a warped product, $X_4\times_W K$, where $X_4$ is AdS$_4$, Mink$_4$, or dS$_4$, and
$K$ is a compact six-dimensional spin manifold. Taking coordinates $x^\m$ on $X_4$ and
$y^m$ on $K$, we write the ten-dimensional string frame metric as
\ba
ds_{string}^2 = e^{\phi/2}\left(e^{2A(y)} \bar{g}_{\m\n}(x) dx^\m dx^\n + e^{-2A(y)/3}\bar{g}_{mn}(y)dy^m dy^n\right).
\ea
The inclusion of a warp factor for the internal metric is purely a matter of convenience, since it can always be absorbed into $\bar{g}_{mn}$. The metric appearing in the brackets corresponds to ten-dimensional Einstein frame. We extract the volume mode from the internal metric by writing $\bar{g}_{mn} = e^{2u}g_{mn}$ for some fiducial metric $g_{mn}$ of fixed volume. To reduce to four-dimensional Einstein frame requires a further rescaling: $\bar{g}_{\m\n} = e^{-6u}g_{\m\n}$. Instead of working with $\f$ and $u$, we will take the combinations
\ba\label{eqn:4ddil}
\vp = \frac{\f}{2} -6u +2A,\qquad \r = \frac{\f}{2} +2u -\frac{2A}{3},
\ea
as the independent dynamical fields, which we will see correspond to the four-dimensional dilaton and volume modulus, respectively. We have absorbed the warp factors into these scalar fields to simplify notation, though we could just as well write $A(y)$ explicitly in what follows. In terms of these quantities, the string frame metric becomes
\ba\label{eqn:4dmetric}
ds_{string}^2 = e^{\vp}g_{\m\n}(x) dx^\m dx^\n + e^{\r}g_{mn}(y) dy^m dy^n.
\ea
To preserve the (maximal) symmetry of the four-dimensional spacetime, we restrict the remaining bosonic fields to only have support along $K$:
\ba
F= \hlf F_{mn}(y)dy^m dy^n,\quad H=\frac{1}{3!}H_{mnp}(y)dy^m dy^n dy^p,\quad \vp = \vp(y),\quad \r = \r(y).
\ea
We only require that these fields be smooth and satisfy the equations of motion (together with the Bianchi identity), but otherwise we leave their form arbitrary. In particular, we make no assumptions about the preservation of supersymmetry or the constraints that would impose on the fields.

\subsection{Four-dimensional ansatz: fermions}\label{ss:fermions}
Having discussed the bosonic fields, we now turn to the fermionic sector. As mentioned before, the gravitino and dilatino play no role in this analysis and so we set them to zero. The gaugino will be important, and we decompose it into its four- and six-dimensional components as
\be\label{eqn:4dspinor}
\chi = e^{-3\vp/4}\left(\chi_4(x)\otimes\chi_6(y) + c.c.\right),
\ee
where we include the prefactor $e^{-3\vp/4}$ to ensure properly normalized kinetic terms in four dimensions, and we are ignoring the higher Kaluza-Klein modes. $\chi_4$ is an adjoint valued, anticommmuting Weyl spinor on $X_4$ of positive chirality, while $\chi_6$ is a gauge singlet, commuting Weyl spinor on $K$, also of positive chirality. The decomposition of gamma matrices is explained in Appendix~\ref{app:conventions}.

If $\chi_4$ is exactly massless, as required for example by supersymmetry, then $\chi_6$ must be a zero mode of the Dirac operator on $K$: $D\!\!\!\!/{}_K \chi_6=0$.  This makes $\chi_6$ covariantly constant, and it is possible to normalize $\chi_6^\dagger\chi_6=1$. However, we will only require the weaker condition
\be\label{eqn:quasi0mode}
D\!\!\!\!/{}_K\chi_6 = O(\a'),
\ee
so that $\chi_6$ is only a ``quasi-zero mode" of the internal Dirac operator. This implies that supersymmetry breaking should be a subleading effect in $\a'$, and $\chi_4$ will be sufficiently light to appear in the low energy effective action. $\chi_6^\dagger\chi_6$ will vary slowly over $K$, and and at best may be normalized to $1+O(\a')$. This minor point, however, will not significantly impact our analysis.

The only fermion bilinear permitted by the maximal symmetry of spacetime is the scalar combination $\langle\chi_4\chi_4\rangle$.
In pure $N=1$ super Yang-Mills theory in four dimensions, such a scalar condensate can develop at low energies where the gauge coupling becomes strong: \footnote{We are setting the axion field to zero, since it will not factor into our analysis.}
\ba\label{eqn:condense}
\left\langle \tr(\chi_4\chi_4)\right\rangle=  c M_{UV}^3 \exp\left(-b g_{YM}^{-2}\right),
\ea
for some numerical constant $c$. We are writing $1/b$ for the one-loop beta function coefficient, and $M_{UV}$ for the UV cutoff scale of the gauge theory. Already in~\cite{Dine:1985rz}\ it was realized that this ansatz for the condensate requires corrections to reconcile the ten and four-dimensional perspectives. We will consider these corrections, and more general forms of the condensate, in Section~\ref{ss:general}.

Following~\cite{Dine:1985rz,Derendinger:1985kk}, we can embed this result into our four-dimensional compactification of the $E_8\times E_8$ heterotic string. We assume that the background field strength $F_{mn}$ leaves the gauge group $G\times E_8$ unbroken, for some $G\subset E_8$,\footnote{For phenomenological applications, we should also require $SU(3)\times SU(2)\times U(1) \subset G$.} and the condensate develops in the hidden $E_8$ (or some non-Abelian subgroup thereof). The UV cutoff scale should then be taken to be the Kaluza-Klein scale of $K$. If $M$ is the typical scale of the metric $g_{mn}$, then the KK scale as measured by $g_{\m\n}$ is
\be\label{eqn:KK}
M_{UV} = e^{(\vp-\r)/2}M.
\ee
In Appendix~\ref{app:reduction}\ we show that the gauge coupling is determined by the four-dimensional dilaton to be
\be
g_{YM}^{2} = e^{\vp}.
\ee
Thus the condensate will manifest itself in the gaugino bilinear three-form:
\ba\label{eqn:condensate}
\tr \chi \tilde\G_{mnp} \chi &= e^{3(\r-\vp)/2}\left(\tr(\chi_4\chi_4)\left(\chi_6^T \g_{mnp}\chi_6\right) +c.c.\right) \\
&\rightarrow c M^3 \exp\left(-b e^{-\vp}\right)\left(\chi_6^T \g_{mnp}\chi_6 +c.c.\right).\non
\ea
On the left hand side we use $\tilde\G$ to denote the ten-dimensional string frame gamma matrices, while those appearing on the right are in four-dimensional Einstein frame. Notice that the conformal factors in~\C{eqn:KK}\ cancel against the normalization of the four-dimensional gaugino~\C{eqn:4dspinor}\ together with the rescaling of $\g_{mnp}$.

Since we are considering both supersymmetric and non-supersymmetric solutions of the heterotic string, we should be cautious in applying~\C{eqn:condense}\ to the latter. However, if supersymmetry is only weakly broken, so that $\chi_4$ remains sufficiently light along the lines discussed above, then it seems reasonable to expect a condensate similar to~\C{eqn:condense}\ to develop.

\section{Maximally symmetric solutions}\label{sec:result}

We carry out the reduction of the ten-dimensional action~\C{eqn:10daction}, in terms of the ansatz of the previous sections, in Appendix~\ref{app:reduction}. We summarize here some of the key results of that procedure. First, $\nab\vp$, $\nab\r$, and $H$ are all $O(\a')$. Second, $H$ and $\tilde R_{+\,6}$ depend on $\r$ but not $\vp$ ($cf.$~\C{eqn:CStilde}\ and~\C{eqn:R+}). On the other hand, $T=H+\frac{a'}{8}\tr(\bar\chi\G\chi)$ depends on $\vp$ and $\chi_6$ though~\C{eqn:condensate}. Finally, the reduced action we wish to study is given by:
\ba\label{eqn:4daction}
S = \int d^4x\sqrt{-g_4}\int d^6y \sqrt{g_6} \left( \cR_4 +\frac{\a'}{4} e^{-\vp}\left|R_4\right|^2 -V +O(\a'^3)\right),
\ea
where the potential $V$ depends only on the internal fields,
\ba\label{eqn:4dpotential}
V=-e^{\vp}&\left[e^{-\r}\left(\cR_6 -|\nabla\rho|^2-\nabla\r\cdot\nabla\vp\right) -\hlf e^{-3\r}\left|T\right|^2 \right. \\ &\left. -\frac{\a'}{4}e^{-2\r}\left(\tr|F|^2-\tr|\tilde R_{6\,+}|^2+ 2c M^3 \exp\left(-b e^{-\vp}\right)\left(\chi_6^T\g^m D_m\chi_6 +c.c.\right)\right)\right]\non.
\ea
As in~\cite{Gukov:2003cy}, our action has a non-standard normalization, but this can easily be corrected by rescaling the spacetime metric: $g_4\rightarrow e^{-\vp_0}g_4$, where $\vp_0$ is the zero-mode of $\vp(y)$.\footnote{Such a rescaling leads to the more conventional looking four-dimensional action
\ba
S = e^{-\vp_0}{\cal V}\int d^4x\sqrt{-g_4}\left(\cR_4 +\frac{\a'}{4}|R_4|^2 - \frac{1}{\cal V}\int d^6y \sqrt{g_6}\ V\right),\non
\ea
where ${\cal V}$ the (fixed) volume of $K$ with respect to the fiducial metric $g_{mn}$, and we have used~\C{eqn:zeromode}.} Doing so will not affect our final result, and since we do not wish to treat $\vp_0$ separately from the rest of $\vp(y)$, we choose to work with the given normalization.

Let us now turn our attention to the equations of motion for the reduced action. Given the $\rho$ dependence of $H$ and $\tilde R_{+\,6}$, the equation of motion for $\r$ is far from simple. Fortunately, we will not require it in our argument. The equation for the dilaton $\vp$ is:
\ba
&-\nab^m\left(e^{\vp-\r}\nab_m\r\right) +\frac{\a'}{4}e^{-\vp}|R_4|^2 +V   \label{eqn:phieom}\\
&\quad =\hlf c b\a'M^3 e^{-2\r}\exp\left(-b e^{-\vp}\right) \left[ \chi_6^T \left(\g^m D_m+\frac{1}{24}e^{-\r}\g^{mnp}T_{mnp}\right)\chi_6 +c.c.\right] +O(\a'^3)\non.
\ea
The terms on the right-hand side of this equation arise from the inhomogeneous $\vp$ dependence of the potential, which come solely from the non-perturbative condensate. However, these terms can be set to zero on-shell by imposing the equation of motion for the internal components of the gaugino $\chi_6$:
\ba
&\left(\g^m D_m + \hlf \g^m \nab_m \left( \vp - 2\r - b e^{-\vp} \right) + \frac{1}{24} e^{-\r} \g^{mnp} T_{mnp} \right)\chi_6 =O(\a'^3). \label{eqn:gauginoeom}
\ea
The ``extra" terms that would appear in~\C{eqn:phieom}, $\chi_6^T\g^m\chi_6 \nab_m(\ldots)$, vanish by a familiar identity for commuting spinors in six dimensions.\footnote{Since $\g^m$ are anti-symmetric we have $\chi^T\g^m\chi =\chi^\a \g^m_{\a\b}\chi^\b = \chi^\b\g^m_{\a\b}\chi^\a = -\chi^\b\g^m_{\b\a}\chi^\a= -\chi^T\g^m\chi$.}
Notice that this equation of motion is consistent with the earlier requirement~\C{eqn:quasi0mode}\ that $\chi_6$ be a zero mode to $O(\a')$.
The final equation of motion we require is that of the four-dimensional metric $g_{\m\n}$:
\ba
&\cR_{\m\n} - \hlf g_{\m\n} \cR_4 + \frac{\a'}{4} e^{-\vp} \left( \cR_{\m\l\s\r} \cR_{\n}{}^{\l\s\r} - \hlf g_{\m\n} |R_4|^2 - 2\nab^\l\nab^\s \cR_{\m\l\n\s}\right) = -\hlf g_{\m\n}V +O(\a'^3). \label{eqn:graveom}
\ea
Note that we do not invoke the Lemma from~\cite{Bergshoeff:1989de}, which states that variations of the action with respect to $\om_+$ vanish on-shell at $O(\a')$, since we are interested in solutions including $O(\a'^2)$ corrections. Thus, \C{eqn:graveom}\ is the full gravitational equation of motion to $O(\a'^3)$.

We now seek solutions to the system of equations~\C{eqn:phieom}-\C{eqn:graveom}\ with a maximally symmetric spacetime. Inserting
\be
\cR_{\m\n\l\s} = \frac{\La}{3}\left(g_{\m\l}g_{\n\s} - g_{\m\s}g_{\n\l}\right)
\ee
into the Einstein equation~\C{eqn:graveom}, we simply obtain
\ba\label{eqn:Lambda}
\La = \hlf V = -\frac{\a'}{6}e^{-\vp}\La^2 +\hlf \nab^m\left(e^{\vp-\r}\nab_m\r\right) +O(\a'^3),
\ea
where in the last step we have imposed~\C{eqn:phieom}\ and~\C{eqn:gauginoeom}. It is interesting to note that the gravitational $\a'$ corrections to the Einstein equations vanish on solutions with maximal symmetry. Integrating~\C{eqn:Lambda}\ over the compact internal space, the divergence term drops out and we are left with
\be\label{eqn:integrated}
\La =-\frac{\a'}{6}e^{-\vp_0}\La^2 + O(\a'^3).
\ee
Only the zero-more of $\vp(y)$ survives in the $O(\a')$ term because
\ba\label{eqn:zeromode}
\int d^6y\sqrt{g_6}e^{-\vp} = e^{-\vp_0}\int d^6y\sqrt{g_6},
\ea
which follows by expanding $e^{-\vp}$ in harmonics. Since~\C{eqn:integrated}\ tells us that $\La$ must be at least $O(\a')$, the only consistent\footnote{We ignore the other solution to~\C{eqn:integrated}, with $\La\sim -1/\a'$, since it is not compatible with a perturbative $\a'$ expansion. See~\cite{Lechtenfeld:2010dr,Chatzistavrakidis:2012qb}\ for more on these solutions.} possibility is
\be
\La = O(\a'^3).
\ee
We conclude that the only solutions satisfying our ansatz, of maximal symmetry together with a gaugino condensate, are flat Minkowski spacetimes. One might worry that the inverse coupling $e^{-\vp_0}\sim1/g_{YM}^2$ appearing in~\C{eqn:integrated}\ could lead to an anomalously large $O(\a')$ correction, thereby invalidating the $\a'$ expansion. However, this is an artifact of our unconventional normalization, discussed below~\C{eqn:4dpotential}. Repeating the analysis with $g_4\rightarrow e^{-\vp_0}g_4$ still results in $\La$ vanishing to $O(\a'^3)$.

\subsection{More general condensates}\label{ss:general}

As we remarked in Section~\ref{ss:fermions}, the ansatz for the gaugino condensate in~\C{eqn:condense}\ is not quite correct for a compactification from ten dimensions. For example, if $K$ is a Calabi-Yau threefold then a combination of H-flux and gaugino condensate results in a supersymmetric theory with superpotential \be\label{eqn:superpot}
W = h +c e^{-bS},
\ee
where Re$(S)=e^{-\vp}$ and $h$ measures the units of H-flux on $K$, roughly given by
\be
h \sim \int_K H\wedge \Omega.
\ee
As pointed out in~\cite{Dine:1985rz}, using the standard K\"ahler potential $K = -\log(S+\bar S) -3\log(T+\bar T)$ for the dilaton and volume modulus  (with $\Re(T)=e^\r$) leads to the potential
\ba\label{eqn:V}
V &= e^K\left[K^{A\bar B}D_A W D_{\bar{B}}\bar W -3|W|^2\right] \,\propto \, e^{\vp-3\r}\left|h+c\left(1+2b e^{-\vp}\right)\exp\left(-be^{-\vp}\right)\right|^2.
\ea
In order to derive this potential from ten dimensions, we must include a factor of $(1+2be^{-\vp})$ in the gaugino condensate. As we will show momentarily, such a modification does not affect our result that $\La$ vanishes to $O(\a'^3)$.

There is no reason to restrict to the formation of a single condensate. In the racetrack scenario (see~\cite{Kaplunovsky:1997cy}\ for a recent review, and references therein) multiple gauge groups are allowed to condense, resulting in a superpotential of the form
\be
W = h+ \sum_a c_a e^{-b_a S}.
\ee
We can easily accommodate this, and countless other possibilities, by replacing the condensate~\C{eqn:condense}\ with an arbitrary non-perturbative function of $g_{YM}^2 =e^\vp$:
\be\label{eqn:condense2}
\left\langle \tr(\chi_4\chi_4)\right\rangle=   M_{UV}^3 f(e^{-\vp}).
\ee
Repeating the analysis of the previous section, we find the dilaton equation,
\ba
&-\nab^m\left(e^{\vp-\r}\nab_m\r\right) +\frac{\a'}{4}e^{-\vp}|R_4|^2 +V   \\
&\quad =-\hlf \a'M^3 f' e^{-2\r} \left[ \chi_6^T \left(\g^m D_m+\frac{1}{24}e^{-\r}\g^{mnp}T_{mnp}\right)\chi_6 +c.c.\right] +O(\a'^3)\non,
\ea
where, once again, the second line vanishes after imposing the internal gaugino equation:
\ba
&\left(\g^m D_m + \hlf \g^m \nab_m \left( \vp - 2\r +\log(f) \right) + \frac{1}{24} e^{-\r} \g^{mnp} T_{mnp} \right)\chi_6 =O(\a'^3).
\ea
The remaining steps are exactly as before, and so we conclude that the generalized ansatz \C{eqn:condense2}\ does not generate a non-zero cosmological constant either.

In fact, we can go one step further: by replacing $f(e^{-\vp})$ by a general function $F(e^{-\vp},e^\r)$, then we still only find Minkowski solutions.
We should point out, however, that it is \emph{not} as far reaching as it may seem. For example, at large volumes, threshold corrections to the (holomorphic) gauge coupling function modify the superpotential~\C{eqn:superpot}\ to
\be
W = h + c e^{-b(S+\beta T)}.
\ee
The $T$ dependence in $W$ breaks the no-scale structure of the original superpotential~\C{eqn:superpot}, so that the scalar potential is no longer a perfect square, as in~\C{eqn:V}. The new terms in $V$ \emph{cannot} be obtained from the ten-dimensional action by simply replacing the condensate with $(polynomial)\times\exp[-b(e^{-\vp}+\beta e^{\r})]$. This makes sense since threshold corrections are a (four-dimensional) one-loop effect, and we are only considering the tree-level action. Thus, threshold corrections  lie outside the scope of our analysis and we cannot rule out the possibility that they can generate a non-trivial cosmological constant. In fact, quite on the contrary, they are known play an essential role in stabilizing moduli at AdS minima in several explicit scenarios~\cite{Gukov:2003cy,Curio:2005ew,Anderson:2011cza,Cicoli:2013rwa}.

\subsection{An all orders conjecture}\label{ss:allord}

The vanishing of the cosmological constant hinges on the manner in which  $\vp$ and $\chi_6$ enter into the potential $V$. If it were not for the non-perturbative contributions from the condensate, $V$ would scale uniformly with $e^{\vp}$. This is a manifestation of the same scaling behaviour in the full tree-level ten-dimensional  string frame action. The non-perturbative corrections, which arise from $\langle \chi_4\chi_4\rangle$, are always accompanied by bilinears of $\chi_6$, which makes perfect sense from a ten-dimensional perspective. This is the reason that the terms on the right-hand side of the dilaton equation~\C{eqn:phieom}\ are proportional to the gaugino equation~\C{eqn:gauginoeom}. Given all of this, it seems reasonable to expect that such cancellations will always hold in the full tree-level effective action. More precisely, we make the following:
\paragraph{Conjecture.}\emph{Suppose we repeat the analysis of this paper starting from the full tree-level heterotic effective action. Let $\Delta\!\!\!\!/\,\chi_6 = Y\chi_6^*$ be the internal gaugino's equation of motion, generalizing~\C{eqn:gauginoeom}\ to all orders in $\a'$, for some operators $\Delta$ and $Y$ that depend on the internal fields (including $\chi_6$). Then, the full equation of motion for $\vp$, generalizing~\C{eqn:phieom}, can be written schematically as:}
\ba\label{eqn:conj}
 V +  \nabla^m(\ldots)_m +\sum_{k,\ell\geq0}\a'^{k+\ell} a_{k,\ell}\nabla^{2k} R_4^{\ell+1}  = C \left(\chi_6^T\left( \Delta\!\!\!\!/\, \chi_6 -Y\chi_6^* \right)+c.c.\right),
\ea
\emph{for some functions $a_{k,\ell}$ and $C$ that depend on the internal fields, but not $g_4$ or $\chi_6$.}

The form of the left-hand side has been argued on general grounds in~\cite{Gautason:2012tb}, in the absence of non-perturbative corrections. Our claim is that the condensate only adds terms proportional to the gaugino's equation of motion. Since $\tr\chi\G_{MNP}\chi$ is the only non-trivial bilinear in ten dimensions built from Majorana-Weyl spinors then, after reducing to four dimensions via~\C{eqn:condensate}, it is immediately clear that our claim holds for all terms in the effective action that depend on $\chi$ but not $D\chi$. To check our claim for derivative couplings will require a careful classification of all possible gaugino-dependent terms in the ten-dimensional effective action, and mapping these to four dimensions. We will not pursue this here, leaving the remaining details of the proof for future work.

Assuming~\C{eqn:conj}\ holds then, once we impose the gaugino equation of motion, the rest of the argument is identical to~\cite{Gautason:2012tb}. Integrating~\C{eqn:conj}\ over the internal space and imposing maximal symmetry ($R_4\sim\La g_4$), the Einstein equations reduce to
\be
\La = \sum_{m,n>0} c_{m,n} \a'^m \La^{n},
\ee
for some constants $c_{m,n}$, obtained by integrating various combinations of internal fields over $K$. The only \emph{perturbative} solution to the above equation is $\La\equiv 0$. Therefore, assuming the validity of our conjectured equations of motion, even in the presence of a gaugino condensate, Minkowski space remains the only perturbative solution with maximal symmetry to the full tree-level heterotic effective action.

\section{Discussion}\label{sec:discussion}

The assumptions required for our argument are rather minimal. In particular, we only assume the validity of an $\a'$ expansion, that the four-dimensional spacetime has maximal symmetry, and that a condensate $\langle \chi_4\chi_4\rangle\sim e^{-1/g^2_{YM}}$ forms. We made no assumptions regarding the form of the remaining fields, or about the preservation of supersymmetry. Our main result is that the only consistent solutions are four-dimensional Minkowski space, at least up to $O(\a'^3)$ corrections. More general forms of the condensate considered do not alter this picture. We have also presented some evidence that this will continue to hold to all orders in an $\a'$ expansion, at string tree-level.

\paragraph{Leading order solutions}
At leading order in $\a'$, the only non-trivial internal field permitted is a generic Ricci-flat metric. The dilaton $\vp$, volume modulus $\r$, and B-field are forced to be constant. This could have been anticipated given the results of~\cite{Witten:1985bz,Witten:1986kg}, where the heterotic equations of motion were first studied in an $\a'$ expansion. There, the leading order solutions were shown to be a Ricci flat manifold with constant $\f$ and vanishing $H$. A non-trivial dilaton or $H$ are only permitted at O$(\a')$. This has also been emphasized more recently (at least for supersymmetric solutions) in~\cite{Melnikov:2014ywa}\ and~\cite{delaOssa:2014msa}. There too, it has been noted that $H$ is $\a'$ suppressed, and the only supersymmetric solutions consistent with an $\a'$ expansion are Calabi-Yau at leading order. The non-K\"ahler solutions of~\cite{Dasgupta:1999ss}, which also have $H\sim O(\a')$, avoid the necessity of Ricci-flatness precisely because they contain string scale cycles and cannot be treated consistently in a perturbative $\a'$ expansion.

\paragraph{Supersymmetry} Since the gauge sector only appears at $O(\a')$ in the heterotic action, the effects of the gaugino field are automatically suppressed. We found that the internal gaugino field $\chi_6$ must be a ``quasi-zero mode" of the internal Dirac operator, so that $D\!\!\!\!/{}_K\chi_6 = O(\a')$. At leading order in $\a'$, this implies the existence of a covariantly constant spinor and reduced holonomy for the (leading order) metric on $K$. Thus $K$ is Calabi-Yau and supersymmetry is preserved up to at least $O(\a')$. This makes sense since, in order for a gaugino condensate to form in the first place, the effects of supersymmetry breaking should be comparatively small. We expect that generic solutions will break supersymmetry at $O(\a')$, however our results are certainly sharpest when supersymmetry is not broken by the background, so that the condensate is reliable and under control.

\paragraph{Fivebranes} Also within the gauge sector, a generic solution for the field strength $F_{mn}$ will contain both instanton and anti-instanton configurations. In the zero size limit, these gauge field configurations become $NS5$/$\overline{NS5}$  wrapping two-cycles ${\cal C}_i\subset K$. In Einstein frame, these solutions are singular and so lie outside the realm of our analysis. Nevertheless, we do not expect any qualitative changes to occur as we smoothly vary an instanton's size from $O(\a')$ down to zero. Thus, we expect that our results will continue to hold in the presence of $NS5$/$\overline{NS5}$. In particular, subject to our assumptions above, the inclusion of $\overline{NS5}$ should not result in a de Sitter spacetime.\footnote{This makes sense since $\overline{NS5}$ have positive tension and so satisfy the strong energy condition.}
 We hope to return to this fascinating problem in the future, and analyze this claim more quantitatively.

\paragraph{Other $g_s$ corrections} 

The leading perturbative corrections to the ten-dimensional effective action are of the form $\a'^3 R^4$. However, unlike their tree-level counterparts these higher-derivative terms do not scale with $e^\vp$, resulting in corrections to the $\vp$ equation of motion~\C{eqn:phieom}\ or~\C{eqn:conj}. It would be interesting to see if these perturbative corrections could result in a non-zero vacuum energy, and specially whether a de Sitter solution is permitted. Unfortunately, before such corrections can be consistently analyzed, the complete one-loop effective action must be computed, and this remains an open challenge. We could also consider other non-perturbative $g_s$ effects, besides gaugino condensation. The main corrections to take into account come from $NS5$-instantons, where Euclidean fivebranes wrap all of $K$. These are related to gauge theory instantons in spacetime.

\paragraph{AdS solutions in heterotic} As explained in the introduction, most heterotic AdS$_4$ solutions rely on additional $\a'$ or $g_s$ effects. However, we were able to identify a few outlying examples,~\cite{deCarlos:2005kh,Frey:2005zz,Held:2010az,Klaput:2012vv}, which generate AdS$_4$ solutions using only the ingredients considered in this analysis: H-flux, bundles, perturbative $\a'$ corrections, and gaugino condensates. The are two likely resolutions to this tension: the first concerns the validity of an $\a'$ expansion in torsional solutions, and the second involves solving the heterotic Bianchi identity. It has been understood since the original work of~\cite{strominger-torsion}\ that heterotic compactifications with H-flux typically contain string-scale cycles. Thus, only checking that the total volume is large does not guarantee that an $\a'$ expansion is valid everywhere on the internal manifold.

The inclusion of H-flux also complicates the Bianchi identity~\C{eqn:bianchi}, making the standard embedding impossible, and finding solutions becomes a non-trivial task. The authors of~\cite{Frey:2005zz}\ attempt to circumvent this problem by modifying the spin connection that appears on the right-hand side of Bianchi.\footnote{It should be possible solve the undeformed Bianchi identity if we allow more exotic phenomena, specifically dilatino condensation~\cite{Manousselis:2005xa}.} However, there is a tight correlation between the connections that appears in the Bianchi identity and in the gravitino's BPS equation, and it is far from obvious that modifying one without the other will yield consistent solutions.\footnote{See~\cite{Becker:2009df,delaOssa:2014msa} and references therein for discussions of this important point.} We hope to return to these examples in future work, to gain a sharper understanding of how this tension is resolved.

\paragraph{AdS solutions in IIB}
On the face of it, our analysis closely mimics the KKLT construction of AdS$_4$ and dS$_4$ solutions in type IIB~\cite{Kachru:2003aw}. In particular, within our heterotic framework we can identify duals of all the necessary ingredients: orientifold planes\footnote{As emphasized in~\cite{Green:2011cn}, type II orientifold planes are dual to $\a'R^2$ couplings in heterotic.}, fluxes, condensates, and potentially even $\overline{D3}$.\footnote{See the above discussion on $\overline{NS5}$-branes.} Thus, it would seem that our results are at odds with these dual constructions, as well. However, duality maps can be subtle; for example, $\a'$ corrections in one frame may correspond to a mixture of $\a'$ and $g_s$ effects in the other (see~\cite{Anguelova:2010ed}, for example). Therefore, we may be missing important $g_s$ corrections on the heterotic side, required for capturing the IIB duals. Clearly, this point deserves further investigation. A more serious objection arises from the very nature of heterotic flux compactifications, which cannot be studied reliably in an $\a'$ expansion. Most likely, the heterotic duals of IIB flux backgrounds contain large curvatures, and so lie outside the regime of this analysis.

%

\section*{Acknowledgements}
It is a pleasure to thank Andrew Frey, Rhiannon Gwyn, and Savdeep Sethi for helpful conversations and suggestions. This work is supported in part by a fellowship from the Pacific Institute for the Mathematical Sciences and by the Natural Sciences and Engineering Research Council of Canada.

\appendix
\section{Conventions}\label{app:conventions}

Here we summarize the conventions we use for indices, forms and spinors. We take our metric to have a ``mostly plus" signature. Capital Roman letters take values $\{0,\ldots,9\}$, lowercase Greek letters take values $\{0,1,2,3\}$, and lowercase Roman letters take values $\{4,\ldots,9\}$. Letters from the beginning of the corresponding alphabet are used for local Lorentz frames, and letters from the middle of the alphabet are used for coordinate bases.
The norm of a rank $p$ tensor is defined by
\be
|T_{(p)}|^2 = \frac{1}{p!} g^{M_1 N_1}\ldots g^{M_pN_p} T_{M_1\ldots M_p}T_{N_1\ldots N_p}.
\ee
We antisymmetrize with an overall normalization of $1/(p!)$, so for example
\be
T_{[M_1\ldots M_p]} =\frac{1}{p!}\sum_\s (-1)^{|\s|} T_{M_{\s(1)}\ldots M_{\s(p)}}
\ee
where $|\s|$ denotes the order of the permutation $\{1,2,\dots, p\}\rightarrow\{\s(1),\s(2),\ldots,\s(p)\}$. In particular, $T_{[MN]} = \hlf(T_{MN} - T_{NM})$.

In a local Lorentz frame, the ten-dimensional gamma matrices satisfy the algebra $\{\G^A,\G^B\}=2\eta^{AB}$. We define $\G^{A_1\cdots A_p} = \G^{[A_1}\cdots \G^{A_p]}$. We use a Majorana basis so that $\G^A$ are real and symmetric, except for $\G^0$ which is antisymmetric. The charge conjugation matrix is given by
$
C =\G^3 \G^5 \G^7\G^9 \G^0,
$
and satisfies
\ba
C^T=-C,\qquad C^2 =-1\!\!1, \qquad C\G^A C^{-1} = -\left(\G^A\right)^T.
\ea
We define $\bar\chi = \chi^T C$, and the above properties imply that $\tr(\bar\chi\G^A\chi)=0$.
The chirality operator in ten dimensions is given by $\G_{(10)} = \G^0\cdots\G^9$.
We decompose the ten-dimensional gamma matrices as
\be
\G^\a = \g^\a\otimes 1\!\!1,\qquad \G^a = \g_{(4)}\otimes \g^a,
\ee
where $\g_{(4)}=-i\g^0\g^1\g^2\g^3$ is the four-dimensional chirality operator, and $\g_{(6)}=i\g^4\cdots\g^9$ is the six-dimensional one. The $\g^\a$ are also real with the same symmetry properties as $\G^\a$, while $\g^a$ are imaginary and antisymmetric. Finally, the Lorentz frame gamma matrices can be expressed in coordinate bases with the vielbein, $\G^M=e^M_A \G^A$. In particular, in a coordinate basis the $\G^M$ transform under conformal transformations of the metric.

\section{Reducing the action}\label{app:reduction}

Reducing the ten-dimensional action~\C{eqn:10daction} on the ansatz of Sections~\ref{sss:ansatz}\ and~\ref{ss:fermions}\ is a straightforward exercise, although there are some slight complications introduced by the conformal factors appearing in the string frame metric~\C{eqn:4dmetric}\ that bear discussion. For example, writing the associated vielbeins as
\be\label{eqn:vielbein}
 e^A_{string} = e^{\f_A/2}e^A,\qquad \f_A = \left\{ \begin{array}{ll} \vp, & A=0,1,2,3, \\ \r, & A=4,5,6,7,8,9
\end{array} \right. ,
\ee
we see there is a shift in the spin connection:
\ba\label{eqn:spinshift}
\om^{AB}_{string} \equiv \tilde\om^{AB} = \om^{AB} -\hlf \left(\nab^A\f_B\right)e^B +\hlf \left(\nab^B\f_A\right)e^A,
\ea
where $\nab^A = e^{AM}\nab_M$. To simplify notation, we will now use tildes to denote string frame quantities, and no tildes for the corresponding ones in four-dimensional Einstein frame. In particular, splitting the local Lorentz indices as $A=\{\a,a\}$, we have
\be
\tilde \om^{\a\b} = \om^{\a\b},\qquad \tilde \om^{\a b} = \hlf e^\a \nab^b\vp,\qquad \tilde \om^{ab} = \om^{ab} + e^{[a} \nab^{b]}\r .
\ee
Furthermore, the conformal transformations~\C{eqn:vielbein}\ imply that $\tilde \G_M =\tilde e_M^A \G_A = e^{\f_A/2}\G_M$.

We can begin at zeroth order in $\a'$, where life is simple and the reduced action is easily determined:
\ba\label{eqn:0thorder}
S_0 =\int d^4x \sqrt{-g_4} \int d^6y  \sqrt{g_6}\left[\cR_4 +e^{\vp-\r}\left(\cR_6 -|\nab\r|^2 - \nab\r\cdot\nab\vp -\hlf e^{-2\r}|dB|^2\right) \right],
\ea
where $\cR_4=g^{\m\n}\cR_{\m\n}$ and $\cR_6=g^{mn}\cR_{mn}$. At higher orders in $\a'$, the effects of~\C{eqn:spinshift}\ appear in nearly every term in the action, which complicates the analysis somewhat. Fortunately, the modification to the gaugino kinetic term vanishes because of the identity $\tr(\bar\chi\G^A\chi)=0$ for Majorana-Weyl spinors. The reduced action in the gauge sector takes the form
\ba\label{eqn:gauge}
S_{gauge} =-\frac{\a'}{2}\int d^4x \sqrt{-g_4}\int d^6y \sqrt{g_6}\left[e^{-\vp}\left(\frac{1}{4}\tr F_{\m\n}F^{\m\n} + (\chi^\dagger_6 \chi_6)\, \tr\bar\chi_D\g^\m D_\m\chi_D \right)\right. \\
\left. +\frac{1}{4} e^{\vp-2\r}\tr F_{mn}F^{mn} +e^{-(\vp+\r)/2}\left(\tr(\chi_4\chi_4)\, \chi_6^T\g^m D_m\chi_6 +c.c.\right)\right],\non
\ea
where $\chi_D$ is the Dirac spinor associated to $\chi_4$.  We have temporarily included the spacetime components $F_{\m\n}$ so that we can identify the four-dimensional gauge coupling,
\be
g_{YM}^{2} = e^{\vp},
\ee
by comparing to the Einstein-Hilbert term~\C{eqn:0thorder}\ and setting the ratio of the gauge and gravitational kinetic terms to $\a'/4$.

On the other hand, the field strength $H$ is modified by~\C{eqn:spinshift} because of the Chern-Simons correction~\C{eqn:H}. The only modification possible to $CS(\om_+)$ is by the addition of an exact three-form, since $dCS(\om)=\tr R\wedge R$ is a topological invariant. A rather lengthy (but straightforward) calculation shows that
\ba\label{eqn:CStilde}
CS(\tilde\om_+) = CS(\om_+) +\sum_A d\left(e^A d\left(\nab^A\f_A\right)\right) = CS(\omega_+) +d\left(e^a d(e^m_a\nab_m\rho)\right),
\ea
where in the last step we imposed $\vp=\vp(y)$. In particular, $H$ is independent of $\vp$.

Finally, we come to the most involved of the couplings, which is $|R_+|^2$.  The full $\vp$ and $\rho$ dependence of these terms are rather complicated, but we can simplify our labour tremendously by the following.

\paragraph{Claim.}\emph{The zeroth order action~\C{eqn:0thorder}\ is extremized by constant $\vp$, $\r$ and $B$. In other words, solutions of the full heterotic effective action have
\ba\label{eqn:claim}
\vp(y) = \vp_0 +O(\a'),\qquad \r(y) = \r_0 +O(\a'),\qquad H(y) =O(\a'),
\ea
for constants $\vp_0$ and $\r_0$.}

An equivalent version of this statement, in terms of the ten-dimensional fields $\f$ and $A$, was derived in~\cite{Gautason:2012tb}\ (see also~\cite{Held:2010az}). We provide a proof of this statement in the subsequent paragraph. Under this simplification, the nonzero components of the curvature tensor are
\ba
\tilde \cR_{+\, \m\n\l\s} &= e^\vp \cR_{\m\n\l\s} +O(\a'^2), \\
\tilde \cR_{+\,\m m\n n} &= -\hlf e^{\vp}g_{\m\n} \left(\nab_m\nab_n\vp +\hlf e^{-\r}H_{mn}{}^r\nab_r\vp\right) +O(\a'^2),\\
\tilde \cR_{+\,mnpq} &= e^{\r}\left(\cR_{mnpq} + g_{q[m}\nab_{n]}\nab_p\r - g_{p[m}\nab_{n]}\nab_q\r\right) +e^{-\r}\nab_{[m}H_{n]pq} + O(\a'^2) \label{eqn:R+}
\ea
where the components of the curvature two-form are related to those of the Riemann tensor in the usual manner:
$
\tilde \cR_{+\, MNPQ} = \tilde e^A_P\, \tilde e^B_Q\, \tilde R_{+\, MNAB}.
$
Note that the mixed index components $\tilde \cR_{+\,\m m\n n}$ will only contribute to the action at $O(\a'^3)$, and so can be ignored.  Furthermore, the purely internal components $\tilde \cR_{+\,mnpq}$ are completely independent of $\vp$ (to this order in $\a'$), which plays an important simplifying role in this paper.

\begin{proof}[Proof of Claim]

From the zeroth order action~\C{eqn:0thorder}, we derive the lowest-order equations of motion for the four-dimensional dilaton $\vp$,  the volume modulus $\r$, and the spacetime metric $g_{\m\n}$:
\ba
-\nab^2\r  &= \cR_6 -2|\nab\r|^2-\hlf e^{-2\r}|dB|^2 +O(\a'),\label{eqn:vp}\\
\nab^2\vp+2\nab^2\r  &= \cR_6 -|\nab\vp|^2+|\nab\r|^2-2\nab\r\cdot\nab\vp -\frac{3}{2}e^{-2\r}|dB|^2 +O(\a'),\label{eqn:rho}\\
\cR_{\m\n} - \hlf g_{\m\n} \cR_4  &= \hlf g_{\m\n} e^{\vp-\r} \left( \cR_6 - |\nab \r|^2 - \nab\r \cdot \nab\vp - \hlf e^{-2\r}|dB|^2\right) +O(\a').\label{eqn:gmunu} 
\ea
We will not have use for the $B$-field equation of motion. The difference of~\C{eqn:vp}\ and~\C{eqn:rho}\ can be written as
\be\label{eqn:diff}
-\nab\cdot\left(e^{\vp-\r}\nab \left(\vp+3\r\right)\right) = e^{\vp-3\r}|dB|^2 +O(\a'),
\ee
which must vanish upon integrating over the  compact internal space. Since the right-hand side is non-negative, this is only possible if
\be\label{eqn:dB}
e^{\vp-3\r}|dB|^2 = O(\a'),
\ee
and so we have
$\nab\cdot\left(e^{\vp-\r}\nab \left(\vp+3\r\right)\right) = O(\a')$.
Multiplying by $(\vp+3\r)$ and integrating by parts, we find
\be
\int d^6y \sqrt{g_6}\  e^{\vp-\r} \left|\nab \left(\vp+3\r\right) \right|^2 = O(\a').
\ee
Since the exponentials can be taken to be $O(1)$, we must have $\nab(\vp+3\r)=O(\a')$ and so the combination $\vp+3\r$ is constant up to $O(\a')$. Therefore demonstrating that either $\vp$ or $\r$ is constant (to this order) is sufficient to prove that they both are.

We now show that $\nab\r=O(\a')$. To do this, we rewrite the dilaton equation~\C{eqn:vp}\ as
\be
-\nab\cdot\left(e^{\vp-\r}\nab\r\right) = e^{\vp-\rho}\left(\cR_6-|\nab\r|^2-\nab\r\cdot\nab\vp\right) +O(\a') = -\hlf \cR_4 +O(\a'),
\ee
where we have simplified using~\C{eqn:dB}, and in the last step used the trace of the Einstein equation~\C{eqn:gmunu}. This vanishes upon integration, but $\cR_4$ is constant on the internal space (and for a maximally symmetric spacetime, it is constant in spacetime as well). So it must be that $\cR_4=O(\a')$, which means $\nab\cdot(e^{\vp-\r}\nab\r)=O(\a')$. Similar to before, we multiply by $\r$ and integrate by parts to find
\be
\int d^6y \sqrt{g_6}\  e^{\vp-\r} \left|\nab \r \right|^2 = O(\a'),
\ee
so that $\nab\r=O(\a')$, as desired. Finally, looking back at~\C{eqn:dB}\ (and assuming $B$ has an expansion in only integer powers of $\a'$) we can  we see that $dB=O(\a')$.
\end{proof}

\ifx\undefined\bysame
\newcommand{\bysame}{\leavevmode\hbox to3em{\hrulefill}\,}
\fi


\begin{thebibliography}{10}

\bibitem{Gibbons:1984kp}
G.~W. Gibbons, {\em {Aspects of Supergravity Theories}}, Three lectures given
  at GIFT Seminar on Theoretical Physics, San Feliu de Guixols, Spain, Jun
  4-11, 1984.

\bibitem{deWit:1986xg}
B.~de~Wit, D.J. Smit, and N.D. Hari~Dass, {\em {Residual Supersymmetry of
  Compactified D=10 Supergravity}}, Nucl.Phys. {\bf B283} (1987) 165.

\bibitem{Maldacena:2000mw}
Juan~Martin Maldacena and Carlos Nunez, {\em {Supergravity description of field
  theories on curved manifolds and a no go theorem}}, Int.J.Mod.Phys. {\bf A16}
  (2001) 822--855, {\tt arXiv:hep-th/0007018} {\tt [hep-th]}.

\bibitem{Gibbons:2003gb}
G.W. Gibbons, {\em {Thoughts on tachyon cosmology}}, Class.Quant.Grav. {\bf 20}
  (2003) S321--S346, {\tt arXiv:hep-th/0301117} {\tt [hep-th]}.

\bibitem{Dasgupta:2014pma}
Keshav Dasgupta, Rhiannon Gwyn, Evan McDonough, Mohammed Mia, and Radu Tatar,
  {\em {de Sitter Vacua in Type IIB String Theory: Classical Solutions and
  Quantum Corrections}}, JHEP {\bf 1407} (2014) 054, {\tt arXiv:1402.5112} {\tt
  [hep-th]}.

\bibitem{Bergshoeff:1989de}
E.~A. Bergshoeff and M.~de~Roo, {\em {The Quartic Effective Action of the
  Heterotic String and Supersymmetry}}, Nucl. Phys. {\bf B328} (1989) 439.

\bibitem{Green:2011cn}
Stephen~R. Green, Emil~J. Martinec, Callum Quigley, and Savdeep Sethi, {\em
  {Constraints on String Cosmology}}, Class.Quant.Grav. {\bf 29} (2012) 075006,
  {\tt arXiv:1110.0545} {\tt [hep-th]}.

\bibitem{Gautason:2012tb}
Fridrik~Freyr Gautason, Daniel Junghans, and Marco Zagermann, {\em {On
  Cosmological Constants from alpha'-Corrections}}, JHEP {\bf 1206} (2012) 029,
  {\tt arXiv:1204.0807} {\tt [hep-th]}.

\bibitem{Curio:2005ew}
Gottfried Curio, Axel Krause, and Dieter Lust, {\em {Moduli stabilization in
  the heterotic/IIB discretuum}}, Fortsch.Phys. {\bf 54} (2006) 225--245, {\tt
  arXiv:hep-th/0502168} {\tt [hep-th]}.

\bibitem{Anguelova:2010qd}
Lilia Anguelova and Callum Quigley, {\em {Quantum Corrections to Heterotic
  Moduli Potentials}}, JHEP {\bf 1102} (2011) 113, {\tt arXiv:1007.5047} {\tt
  [hep-th]}.

\bibitem{Cicoli:2013rwa}
Michele Cicoli, Senarath de~Alwis, and Alexander Westphal, {\em {Heterotic
  Moduli Stabilisation}}, JHEP {\bf 1310} (2013) 199, {\tt arXiv:1304.1809}
  {\tt [hep-th]}.

\bibitem{Kutasov:2015eba}
David Kutasov, Travis Maxfield, Ilarion Melnikov, and Savdeep Sethi, {\em
  {Constraining de Sitter Space in String Theory}}, {\tt arXiv:1504.00056} {\tt
  [hep-th]}.

\bibitem{Dine:1985rz}
Michael Dine, R.~Rohm, N.~Seiberg, and Edward Witten, {\em {Gluino Condensation
  in Superstring Models}}, Phys.Lett. {\bf B156} (1985) 55.

\bibitem{Derendinger:1985kk}
J.P. Derendinger, Luis~E. Ibanez, and Hans~Peter Nilles, {\em {On the
  Low-Energy d = 4, N=1 Supergravity Theory Extracted from the d = 10, N=1
  Superstring}}, Phys.Lett. {\bf B155} (1985) 65.

\bibitem{Gukov:2003cy}
Sergei Gukov, Shamit Kachru, Xiao Liu, and Liam McAllister, {\em {Heterotic
  moduli stabilization with fractional Chern-Simons invariants}}, Phys.Rev.
  {\bf D69} (2004) 086008, {\tt arXiv:hep-th/0310159} {\tt [hep-th]}.

\bibitem{Anderson:2011cza}
Lara~B. Anderson, James Gray, Andre Lukas, and Burt Ovrut, {\em {Stabilizing
  All Geometric Moduli in Heterotic Calabi-Yau Vacua}}, Phys.Rev. {\bf D83}
  (2011) 106011, {\tt arXiv:1102.0011} {\tt [hep-th]}.

\bibitem{Buchbinder:2003pi}
Evgeny~I. Buchbinder and Burt~A. Ovrut, {\em {Vacuum stability in heterotic M
  theory}}, Phys.Rev. {\bf D69} (2004) 086010, {\tt arXiv:hep-th/0310112} {\tt
  [hep-th]}.

\bibitem{Becker:2004gw}
Melanie Becker, Gottfried Curio, and Axel Krause, {\em {De Sitter vacua from
  heterotic M theory}}, Nucl.Phys. {\bf B693} (2004) 223--260, {\tt
  arXiv:hep-th/0403027} {\tt [hep-th]}.

\bibitem{Buchbinder:2004nt}
Evgeny~I. Buchbinder, {\em {Five-brane dynamics and inflation in heterotic
  M-theory}}, Nucl.Phys. {\bf B711} (2005) 314--344, {\tt arXiv:hep-th/0411062}
  {\tt [hep-th]}.

\bibitem{Gray:2007qy}
James Gray, Andre Lukas, and Burt Ovrut, {\em {Flux, gaugino condensation and
  anti-branes in heterotic M-theory}}, Phys.Rev. {\bf D76} (2007) 126012, {\tt
  arXiv:0709.2914} {\tt [hep-th]}.

\bibitem{Serone:2007sv}
Marco Serone and Alexander Westphal, {\em {Moduli Stabilization in Meta-Stable
  Heterotic Supergravity Vacua}}, JHEP {\bf 0708} (2007) 080, {\tt
  arXiv:0707.0497} {\tt [hep-th]}.

\bibitem{deCarlos:2005kh}
Beatriz de~Carlos, Sebastien Gurrieri, Andre Lukas, and Andrei Micu, {\em
  {Moduli stabilisation in heterotic string compactifications}}, JHEP {\bf
  0603} (2006) 005, {\tt arXiv:hep-th/0507173} {\tt [hep-th]}.

\bibitem{Frey:2005zz}
Andrew~R. Frey and Matthew Lippert, {\em {AdS strings with torsion: Non-complex
  heterotic compactifications}}, Phys.Rev. {\bf D72} (2005) 126001, {\tt
  arXiv:hep-th/0507202} {\tt [hep-th]}.

\bibitem{Held:2010az}
Johannes Held, Dieter Lust, Fernando Marchesano, and Luca Martucci, {\em {DWSB
  in heterotic flux compactifications}}, JHEP {\bf 1006} (2010) 090, {\tt
  arXiv:1004.0867} {\tt [hep-th]}.

\bibitem{Klaput:2012vv}
Michael Klaput, Andre Lukas, Cyril Matti, and Eirik~E. Svanes, {\em {Moduli
  Stabilising in Heterotic Nearly K\'ahler Compactifications}}, JHEP {\bf 1301}
  (2013) 015, {\tt arXiv:1210.5933} {\tt [hep-th]}.

\bibitem{Lechtenfeld:2010dr}
Olaf Lechtenfeld, Christoph Nolle, and Alexander~D. Popov, {\em {Heterotic
  compactifications on nearly Kahler manifolds}}, JHEP {\bf 1009} (2010) 074,
  {\tt arXiv:1007.0236} {\tt [hep-th]}.

\bibitem{Chatzistavrakidis:2012qb}
Athanasios Chatzistavrakidis, Olaf Lechtenfeld, and Alexander~D. Popov, {\em
  {Nearly K\'ahler heterotic compactifications with fermion condensates}}, JHEP
  {\bf 1204} (2012) 114, {\tt arXiv:1202.1278} {\tt [hep-th]}.

\bibitem{Kaplunovsky:1997cy}
Vadim Kaplunovsky and Jan Louis, {\em {Phenomenological aspects of F theory}},
  Phys.Lett. {\bf B417} (1998) 45--49, {\tt arXiv:hep-th/9708049} {\tt
  [hep-th]}.

\bibitem{Witten:1985bz}
Edward Witten, {\em {New Issues in Manifolds of SU(3) Holonomy}}, Nucl.Phys.
  {\bf B268} (1986) 79.

\bibitem{Witten:1986kg}
Louis Witten and Edward Witten, {\em {Large Radius Expansion of Superstring
  Compactifications}}, Nucl. Phys. {\bf B281} (1987) 109.

\bibitem{Melnikov:2014ywa}
Ilarion~V. Melnikov, Ruben Minasian, and Savdeep Sethi, {\em {Heterotic fluxes
  and supersymmetry}}, JHEP {\bf 1406} (2014) 174, {\tt arXiv:1403.4298} {\tt
  [hep-th]}.

\bibitem{delaOssa:2014msa}
Xenia de~la Ossa and Eirik~Eik Svanes, {\em {Connections, Field Redefinitions
  and Heterotic Supergravity}}, JHEP {\bf 1412} (2014) 008, {\tt
  arXiv:1409.3347} {\tt [hep-th]}.

\bibitem{Dasgupta:1999ss}
K.~Dasgupta, G.~Rajesh, and S.~Sethi, {\em M theory, orientifolds and
  {G}-flux}, JHEP {\bf 08} (1999) 023, {\tt arXiv:hep-th/9908088}.

\bibitem{strominger-torsion}
A.~Strominger, {\em Superstrings with torsion}, Nucl. Phys. B. {\bf 274} (1986)
  253.

\bibitem{Manousselis:2005xa}
Pantelis Manousselis, Nikolaos Prezas, and George Zoupanos, {\em
  {Supersymmetric compactifications of heterotic strings with fluxes and
  condensates}}, Nucl.Phys. {\bf B739} (2006) 85--105, {\tt
  arXiv:hep-th/0511122} {\tt [hep-th]}.

\bibitem{Becker:2009df}
K.~Becker and S.~Sethi, {\em {Torsional Heterotic Geometries}}, Nucl. Phys. B
  {\bf 820} (2009) 1--31, {\tt arXiv:0903.3769} {\tt [hep-th]}.

\bibitem{Kachru:2003aw}
Shamit Kachru, Renata Kallosh, Andrei~D. Linde, and Sandip~P. Trivedi, {\em {De
  Sitter vacua in string theory}}, Phys.Rev. {\bf D68} (2003) 046005, {\tt
  arXiv:hep-th/0301240} {\tt [hep-th]}.

\bibitem{Anguelova:2010ed}
Lilia Anguelova, Callum Quigley, and Savdeep Sethi, {\em {The Leading Quantum
  Corrections to Stringy Kahler Potentials}}, JHEP {\bf 1010} (2010) 065, {\tt
  arXiv:1007.4793} {\tt [hep-th]}.

\end{thebibliography}

\end{document}